# Dynamic Phasor Modeling of Single-Phase Grid-Forming Converters

Wenjia Si, *Student Member, IEEE*, Chenming Liu, *Student Member, IEEE,* Steven Liu, *Member, IEEE*, Hongchang Li, *Senior Member, IEEE,* Chenghui Zhang, *Fellow, IEEE,* and Jingyang Fang, *Senior Member, IEEE*

*Abstract*—In modern power systems, grid-forming power converters (GFMCs) have emerged as an enabling technology. However, the modeling of single-phase GFMCs faces new challenges. In particular, the nonlinear orthogonal signal generation unit, crucial for power measurement, still lacks an accurate model. To overcome the challenges, this letter proposes a dynamic phasor model of single-phase GFMCs. Moreover, we linearize the proposed model and perform stability analysis, which confirm that the proposed model is more accurate than existing models. Experimental results validate the improved accuracy of the proposed dynamic phasor model.

## I. INTRODUCTION

Modern power systems are transitioning from centralized fossil-fuel-based generation to distributed renewable energy generation, where renewable energies are coupled to the grid through grid-tied converters. With the capability of active grid support, grid-forming converters (GFMC) emerge as an enabling technology. Since distributed renewable energy sources often feature small sizes, there is a growing demand for single-phase GFMCs [1].

Recently, the modeling of GFMCs has become a research hotspot, which lays a foundation for the stability analysis, control, and optimization of GFMCs. Impedance and state-space modeling are two standard GFMC modeling methodologies, which are based on classical and modern control theory, respectively [2]–[3]. Moreover, reference [4] builds a dynamic phasor model of three-phase GFMCs, which takes into account the dynamics of transmission lines. However, the derivation process of this model is not rigorous, and it cannot accurately describe the dynamics of single-phase GFMCs.

In terms of power fluctuations, single-phase GFMCs differ from three-phase GFMCs. The instantaneous power fluctuations of three phases will cancel out and lump into a flattened sum power of three-phase GFMCs. Through *abc*/*dq*0 frame transformations and power measurement, we can obtain average reactive/active power without any time delay. In contrast, single-phase GFMCs have inherent power fluctuations and require nonlinear orthogonal signal generation units to measure the average power [1]. However, the dynamics of orthogonal signal generation units have been ignored, leading to the inaccuracy of existing models. As a result, the instability of GFMCs will be hidden.

To address the challenge of the inaccurate modeling of single-phase GFMCs, this letter proposes a dynamic phasor model. The contributions of this letter are threefold:

1) Proposing an accurate dynamic phasor model of single-phase GFMCs;
2) Deriving the linearized dynamic phasor model;
3) Performing stability analysis using the proposed model.

The rest of this letter is organized as follows. Section II introduces the schematic and control structure of single-phase GFMCs. Section III proposes a dynamic phasor model for single-phase GFMCs and linearizes it. Additionally, the section performs stability analysis and compares the proposed model with existing models. Section IV provides experimental results for validation purposes. Section V concludes this letter.

## II. SYSTEM CONFIGURATION AND CONTROL STRUCTURE

Fig. 1 shows a schematic of a single-phase grid-forming converter. The grid is modelled as a serial connection of the inductor $L_s$, resistor $R_s$, and voltage source $v_s$. The grid-forming converter is coupled to the grid via an *LCL* filter (including $L_{gi}$, $C_{gf}$, and $L_{gg}$). $v_{gf}$ and $i_{gi}$ represent the converter-side voltage and current, respectively. $v_{gg}$ is the voltage at the point of common coupling (PCC). $i_{gg}$ stands for the grid-injected current. $v_{gdc}$ represents the dc bus voltage.

Moreover, Fig. 1 illustrates the control loops and measurement unit block diagrams of the GMFC system. Generally, GFMCs feature three cascaded control loops—inner current $G_i(s)$, middle voltage $G_v(s)$, and outer power $G_p(s)/G_q(s)$ control loops, where *s* denotes the Laplacian complex frequency variable. Due to the fast and precise voltage and current control, the dynamics of voltage and current control loops can be ignored [4]. In addition, the subscript ref represents reference variables. $\theta_g/v_{gf\_am}$ denotes the phase angle/voltage amplitude generated by the active/reactive power controller. $p_g/q_g$ stands for the active/reactive power injected into the grid. As per [5], $p_g$ and $q_g$ can be calculated via the following equations:

$$\begin{bmatrix} p_g(t) \\ q_g(t) \end{bmatrix} = \frac{1}{2}\begin{bmatrix} v_{gf}(t) & v_{gf}(t-T_0/4) \\ v_{gf}(t-T_0/4) & -v_{gf}(t) \end{bmatrix}\begin{bmatrix} i_{gg}(t) \\ i_{gg}(t-T_0/4) \end{bmatrix}, \quad (1)$$

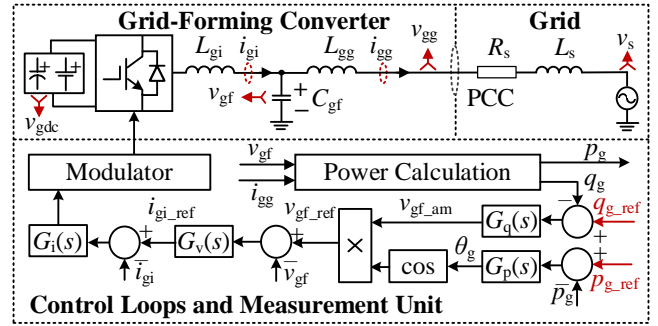

**Fig. 1.** Schematics and control block diagram of grid-forming converters.

where the orthogonal signals are generated by $T_0/4$ (where $T_0 = 1/f_0$, and $f_0$ is the line frequency) time shifting.

## III. Dynamic Phasor Model of Single-Phase GFMCs

This section proposes a dynamic phasor model of single-phase GFMCs.

### A. Dynamic Phasor Modeling

The $n$th dynamic phasor at time $t$, $\langle \mathbf{x} \rangle_n(t)$, is defined as

$$\langle \mathbf{x} \rangle_n(t) = \frac{1}{T_0} \int_{t-\frac{T_0}{2}}^{t+\frac{T_0}{2}} x(\tau) e^{-jn\omega_0 \tau} d\tau, \quad (2)$$

where $\omega_0$ is the fundamental angular frequency [6]. The inverse transform is expressed as

$$x(\tau) = \sum_{n=-\infty}^{+\infty} \langle \mathbf{x} \rangle_n(t) e^{jn\omega_0 \tau}, \quad t - T_0/2 \leq \tau \leq t + T_0/2. \quad (3)$$

Next, the dynamic phasor models of individual units in Fig. 1 are derived as follows:

1) The grid voltage $v_s$ can be expressed in the time domain as

$$v_s(t) = V_s \cos(\omega_0 t), \quad (4)$$

where $V_s$ is the voltage amplitude of $v_s$. Per (2), the grid voltage model can be derived as

$$\langle \mathbf{v_s} \rangle_k(t) = V_s/2, k = \pm 1 \quad (5)$$

where $k = \pm 1$ denotes the 1st dynamic phasor.

2) The GFMC voltage $v_{gf}$ can be expressed in the time domain as

$$v_{gf}(t) = v_{gf\_am}(t) \cos \theta_g(t) = v_{gf\_am}(t) \cos[\omega_0 t + \varphi_g(t)], \quad (6)$$

where $v_{gf\_am}$ (i.e., the voltage amplitude) and $\varphi_g$ (i.e., the phase difference between $v_{gf}$ and $v_s$) are slowly varying, thereby they can be treated as constants in a fundamental period. Per (2), the model of the GFMC voltage can be derived as

$$\langle \mathbf{v_{gf}} \rangle_k(t) = v_{gf\_am}(t) e^{jk\varphi_g(t)}/2, k = \pm 1. \quad (7)$$

3) Per [6], the relationship between $v_L$ (i.e., the voltage of transmission line inductors $L_{gs} = L_{gg} + L_{gs}$) and $i_{gg}$ can be expressed by

$$\langle \mathbf{v_L} \rangle_k(t) = L_{gs} \frac{d\langle \mathbf{i_{gg}} \rangle_k(t)}{dt} + jk\omega_0 L_{gs} \langle \mathbf{i_{gg}} \rangle_k(t), k = \pm 1, \quad (8)$$

which incorporates the inductor dynamics. Moreover, the resistor voltage $v_R$ can be expressed as

$$\langle \mathbf{v_R} \rangle_k(t) = R_{gs} \langle \mathbf{i_{gg}} \rangle_k(t), k = \pm 1 \quad (9)$$

4) The model of the grid-tied current $i_{gg}$ can be derived from the Kirchhoff's voltage law (KVL) as

$$L_{gs} \frac{d\langle \mathbf{i_{gg}} \rangle_k(t)}{dt} + jk\omega_0 L_{gs} \langle \mathbf{i_{gg}} \rangle_k(t) + R_{gs} \langle \mathbf{i_{gg}} \rangle_k(t) = \langle \mathbf{v_{gf}} \rangle_k(t) - \langle \mathbf{v_s} \rangle_k(t), k = \pm 1. \quad (10)$$

5) Per (1), the power measurement part involves the product of two original signals, the product of two time-shifting signals, and the product of one time-shifting signal and one original signal, which can be derived from (2) and (3) as

$$\langle \mathbf{xy} \rangle_n(t) = \sum_{k=-\infty}^{+\infty} \langle \mathbf{x} \rangle_{n-k}(t) \langle \mathbf{y} \rangle_k(t), \quad (11)$$

$$\langle \mathbf{xy} \rangle_n(t-t_0) = \sum_{k=-\infty}^{+\infty} \langle \mathbf{x} \rangle_{n-k}(t-t_0) \langle \mathbf{y} \rangle_k(t-t_0), \quad (12)$$

$$\langle \mathbf{x^d y} \rangle_n(t-t_0) = \sum_{k=-\infty}^{+\infty} \langle \mathbf{x} \rangle_{n-k}(t-t_0) \langle \mathbf{y} \rangle_k(t-t_0) e^{jk\omega t_0}, \quad (13)$$

respectively. $t_0$ is the shifted time, which equals $T_0/4$ in this letter. In (13), the superscript d denotes the time-shifting signal. Per (1), (11)–(13), active and reactive power can be derived as

$$\langle p_g \rangle_0(t) = \frac{1}{2} \sum_{k=-1,1} [\langle \mathbf{v_{gf}} \rangle_k(t) \langle \mathbf{i_{gg}} \rangle_{-k}(t) + \langle \mathbf{v_{gf}} \rangle_k(t-t_0) \langle \mathbf{i_{gg}} \rangle_{-k}(t-t_0)], \quad (14)$$

$$\langle q_g \rangle_0(t) = j \sum_{k=-1,1} -k \langle \mathbf{v_{gf}} \rangle_k(t-t_0) \langle \mathbf{i_{gg}} \rangle_{-k}(t-t_0). \quad (15)$$

where the 0th dynamic phasors refer to the average power.

### B. Model Linearization

Mreover, we perform model linearization by replacing the derived dynamic phasors with the sum of their steady-state values $\langle \mathbf{X} \rangle_n$ (i.e., uppercase letters) and perturbation values $\Delta \langle \mathbf{x} \rangle_n(t)$ (i.e., prefixed with $\Delta$). Through the linearization and Laplace transformation, the small signal expressions of (14) and (15) in the complex frequency domain are derived as

$$\Delta \langle p_g \rangle_0(s) \approx \frac{1+e^{-st_0}}{2} \sum_{k=-1,1} [\Delta \langle \mathbf{v_{gf}} \rangle_k(s) \langle \mathbf{I_{gg}} \rangle_{-k} + \langle \mathbf{V_{gf}} \rangle_k \Delta \langle \mathbf{i_{gg}} \rangle_{-k}(s)], \quad (16)$$

$$\Delta \langle q_g \rangle_0(s) \approx j e^{-st_0} \sum_{k=-1,1} -k[\Delta \langle \mathbf{v_{gf}} \rangle_k(s) \langle \mathbf{I_{gg}} \rangle_{-k} + \langle \mathbf{V_{gf}} \rangle_k \Delta \langle \mathbf{i_{gg}} \rangle_{-k}(s)], \quad (17)$$

whose elements are derived from (7) and (10) as

$$\begin{cases} \Delta \langle \mathbf{v_{gf}} \rangle_k(s) \approx [jkV_{gf\_am} e^{jk\Phi_g} \Delta \varphi_g(s) + \Delta v_{gf\_am}(s) e^{jk\Phi_g}]/2 \\ \Delta \langle \mathbf{i_{gg}} \rangle_k(s) \approx \frac{jkV_{gf\_am} e^{jk\Phi_g} \Delta \varphi_g(s) + \Delta v_{gf\_am}(s) e^{jk\Phi_g}}{2(L_{gs}s + jk\omega_0 L_{gs} + R_{gs})} \\ \langle \mathbf{V_{gf}} \rangle_k = V_{gf\_am} e^{jk\Phi_g}/2 \\ \langle \mathbf{I_{gg}} \rangle_k = \frac{1}{2} \frac{V_{gf\_am} e^{jk\Phi_g} - V_s}{jk\omega_0 L_{gs} + R_{gs}} \end{cases}, k = \pm 1. \quad (18)$$

Substitution of (18) into (16) and (17) yields

$$\begin{cases} \Delta \langle p_g \rangle_0(s) \approx k_{p\varphi}(s) \Delta \varphi_g(s) + k_{pv}(s) \Delta v_{gf\_am}(s) \\ \Delta \langle q_g \rangle_0(s) \approx k_{q\varphi}(s) \Delta \varphi_g(s) + k_{qv}(s) \Delta v_{gf\_am}(s) \end{cases}, \quad (19)$$

where $k_{p\varphi}(s)$, $k_{pv}(s)$, $k_{q\varphi}(s)$, and $k_{qv}(s)$ are expressed as

$$\begin{cases} k_{p\varphi}(s) = k_1(s)[\frac{-V_{gf\_am}^2 X_{gs} + V_{gf\_am} V_s (\cos\Phi_g X_{gs} + \sin\Phi_g R_{gs})}{X_{gs}^2 + R_{gs}^2} + \frac{V_{gf\_am}^2 X_{gs}}{(L_{gs}s + R_{gs})^2 + X_{gs}^2}] \\ k_{pv}(s) = k_1(s)[\frac{V_s \sin\Phi_g \omega_0 L_{gs} + (V_{gf\_am} - V_s \cos\Phi_g) R_{gs}}{X_{gs}^2 + R_{gs}^2} + \frac{V_{gf\_am}(L_{gs}s + R_{gs})}{(L_{gs}s + R_{gs})^2 + X_{gs}^2}] \\ k_{q\varphi}(s) = k_2(s)[\frac{V_{gf\_am}^2 R_{gs} + V_{gf\_am} V_s (\sin\Phi_g X_{gs} - \cos\Phi_g R_{gs})}{X_{gs}^2 + R_{gs}^2} - \frac{V_{gf\_am}^2 (L_{gs}s + R_{gs})}{(R_{gs} + L_{gs}s)^2 + X_{gs}^2}] \\ k_{qv}(s) = k_2(s)[\frac{V_{gf\_am} X_{gs} - V_s (\cos\Phi_g X_{gs} + \sin\Phi_g R_{gs})}{X_{gs}^2 + R_{gs}^2} + \frac{V_{gf\_am} X_{gs}}{(R_{gs} + L_{gs}s)^2 + X_{gs}^2}] \end{cases}, \quad (20)$$

where $X_{gs} = \omega_0 L_{gs}$, $k_1(s) = (1+e^{-st_0})/4$, and $k_2(s) = e^{-st_0}/2$. The dynamics of the orthogonal signal generation unit are modelled. Further, the time delay $e^{-st_0}$ can be simplified as $(-t_0 s + 2)/(t_0 s + 2)$ by the Pade approximation. Moreover, the small signal model of the active power controller (APC) and reactive power controller (RPC) with the virtual synchronous machine control can be written as [3]

$$\begin{cases} \Delta \varphi_g(s) = \frac{\omega_0}{S_0(2H_g s + D_g)s}[\Delta p_{g\_ref}(s) - \Delta p_g(s)] = \frac{-\Delta p_g(s)\omega_0}{S_0(2H_g s + D_g)s} \\ \Delta v_{gf\_am}(s) = k_{q1}[\Delta q_{g\_ref}(s) - \Delta q_g(s)] = -k_{q1}\Delta q_g(s), k_{q1} = V_0 k_q/S_0 \end{cases}, \quad (21)$$

where $H_g$ and $D_g$ stand for the inertia and damping coefficients, respectively. $S_0$ and $V_0$ represent the rated power and voltage, respectively. $k_q$ denotes the reactive power droop gain.

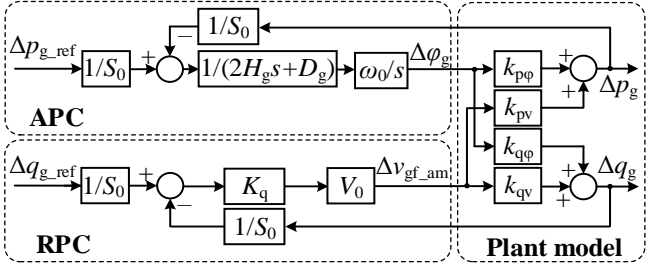

Fig. 2. Small-signal model of single-phase GFMCs.

Combining the small-signal models of the plant and the controllers, we obtain the small-signal model shown in Fig. 2. According to (19) and (21), the characteristic equation of the proposed model can be expressed as

$$[1+k_{q1}k_{qv}(s)]S_0(2H_g s^2 + D_g s) + \omega_0 k_{p\varphi}(s)$$
$$+k_{q1}\omega_0[k_{qv}(s)k_{p\varphi}(s) - k_{pv}(s)k_{q\varphi}(s)] = 0. \quad (22)$$

### C. Stability Analysis Based on the Proposed Model

As shown in Fig. 3, we change the damping coefficient $D_g$ from 4 to 30 to perform stability analysis, and the other parameters are listed in Table I. In particular, when $D_g = 4$, the proposed model shows that some eigenvalues [as calculated from (22)] are located in the right half plane, indicating system instability. However, the model proposed in [3], which does not consider the dynamics of various units of the system, and

Table I. System and control parameters.

| Descriptions | Symbols | Values |
|---|---|---|
| Rated/Grid voltage | $V_0/V_s$ | 110 Vrms |
| Rated power/frequency | $S_0/f_0$ | 1000 VA/50 Hz |
| Active/Reactive power reference | $p_{g\_ref}/q_{g\_ref}$ | 250 W/0 Var |
| Line inductance/resistance | $L_{gs}/R_s$ | 8 mH/0.3 Ω |
| Reactive power droop gain | $k_q$ | 0.05 |
| Inertia/Damping coefficient | $H_g/D_g$ | 5 s/10 |

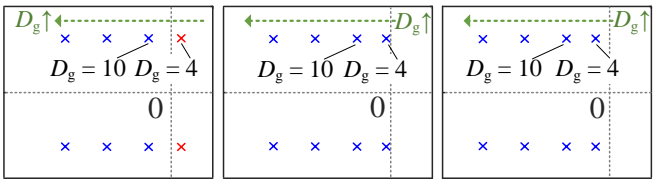

(a) proposed model (b) model in [4] (c) model in [3]

Fig. 3. Dominant eigenvalues with the change of $D_g$ from 4 to 30.

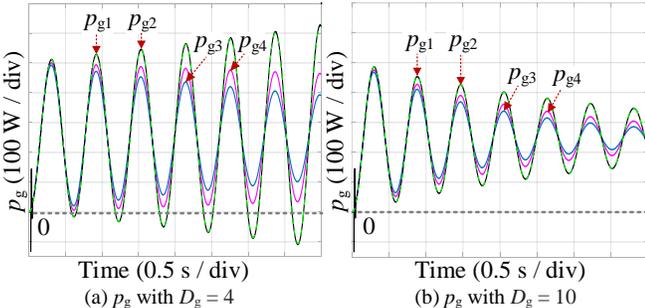

(a) $p_g$ with $D_g = 4$ (b) $p_g$ with $D_g = 10$

Fig. 4. Simulated converter output active power $p_g$ with $D_g = 4$ and 10 in the cases of the detailed system model (i.e., $p_{g1}$), the proposed model (i.e., $p_{g2}$), the model in [3] (i.e., $p_{g3}$), and the model in [4] (i.e., $p_{g4}$).

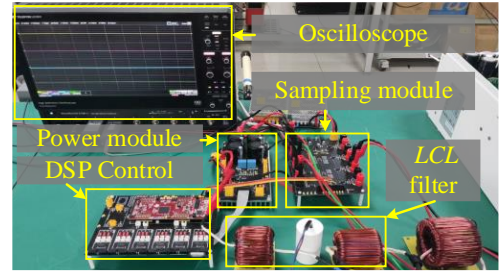

Fig. 5. A photo of the experimental setup.

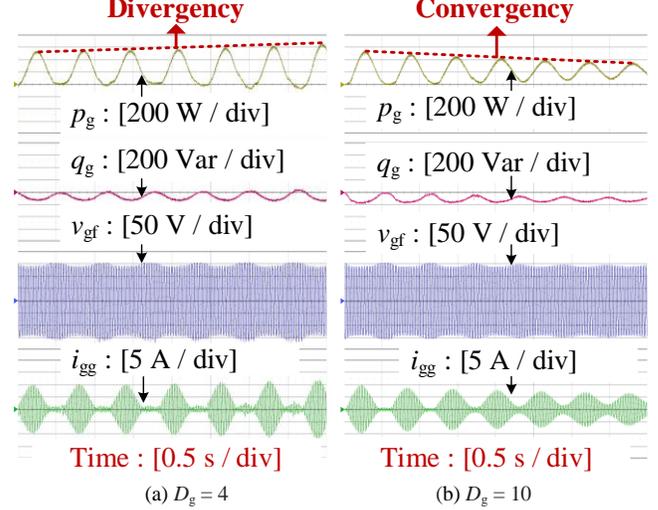

(a) $D_g = 4$ (b) $D_g = 10$

Fig. 6. Experimental results of the single-phase GFMC with $D_g = 4$ and 10.

[4], which only considers the dynamics of the transmission line, yield stable systems.

For verifications, simulation results for different models with $D_g = 4$ and $D_g = 10$ are provided in Fig. 4. As for the detailed system model simulation results $p_{g1}$, the system is stable when $D_g = 10$, but unstable when $D_g = 4$. These results are consistent with the proposed dynamic phasor model simulation results (i.e., $p_{g2}$), showing that the proposed model can well predict the system stability. While, $p_{g3}$ (i.e., the simulation results of the model proposed in [3]) and $p_{g4}$ (i.e., the simulation results of the model proposed in [4]) are inaccurate.

### IV. EXPERIMENTAL VERIFICATIONS

Fig. 5 presents a photo of the experimental setup. The single-phase GFM comprises a control module, a power module, and a sampling module, etc. The GFMC was fed by a dc power supply and connected to the grid emulated by an ac power supply via an *LCL* filter. An oscilloscope captured all the waveforms. All the parameters are listed in Table I.

Fig. 6 depicts the experimental results of the single-phase GFMC with two damping coefficients $D_g = 4$ and 10, respectively. Experimental waveforms include the active power $p_g$, the reactive power $q_g$, the converter voltage $v_{gf}$, and the converter current $i_{gg}$. As shown, the single-phase GFMC operates unstably (indicated by the diverging experimental waveforms) with $D_g = 4$, while operating stably (indicated by the converging experimental waveforms) with $D_g = 10$. The experimental results are consistent with theoretical analysis, validating the accuracy of the proposed model.

## V. Conclusion

This letter builds an accurate dynamic phasor model of single-phase GFMCs. The proposed model can capture the dynamics of important system elements, particularly for the orthogonal signal generation unit. Moreover, we derive a linearized model for small-signal stability analysis. The stability analysis results verify that the proposed dynamic phasor model can accurately predict the system stability, while other existing models fail to do so. Experimental results validate the accuracy of the proposed model.